\pdfoutput=1
\documentclass[12pt,a4paper]{article}
\usepackage{ifthen}
\newboolean{pdflatex}
\setboolean{pdflatex}{true}
\newboolean{articletitles}
\setboolean{articletitles}{true}
\newboolean{uprightparticles}
\setboolean{uprightparticles}{false}
\newboolean{inbibliography}
\setboolean{inbibliography}{false}

\usepackage{microtype}
\usepackage{lineno}  
\usepackage{xspace} 
\usepackage{caption} 

\usepackage{graphicx}  
\usepackage{color}
\usepackage{colortbl}
\graphicspath{{./figs/}} 

\usepackage{amsmath} 
\usepackage{amssymb}
\usepackage{amsfonts}
\usepackage{upgreek} 

\newcommand*\patchAmsMathEnvironmentForLineno[1]{%
\expandafter\let\csname old#1\expandafter\endcsname\csname #1\endcsname
\expandafter\let\csname oldend#1\expandafter\endcsname\csname
end#1\endcsname
 \renewenvironment{#1}%
   {\linenomath\csname old#1\endcsname}%
   {\csname oldend#1\endcsname\endlinenomath}%
}
\newcommand*\patchBothAmsMathEnvironmentsForLineno[1]{%
  \patchAmsMathEnvironmentForLineno{#1}%
  \patchAmsMathEnvironmentForLineno{#1*}%
}
\AtBeginDocument{%
\patchBothAmsMathEnvironmentsForLineno{equation}%
\patchBothAmsMathEnvironmentsForLineno{align}%
\patchBothAmsMathEnvironmentsForLineno{flalign}%
\patchBothAmsMathEnvironmentsForLineno{alignat}%
\patchBothAmsMathEnvironmentsForLineno{gather}%
\patchBothAmsMathEnvironmentsForLineno{multline}%
\patchBothAmsMathEnvironmentsForLineno{eqnarray}%
}

\usepackage{hyperref}    
\usepackage[all]{hypcap} 

\def\lhcb {\mbox{LHCb}\xspace}

\def\belle  {\mbox{Belle}\xspace}

\def\cdf    {\mbox{CDF}\xspace}

\def\MagUp {\mbox{\em Mag\kern -0.05em Up}\xspace}

\ifthenelse{\boolean{uprightparticles}}%
{

 \def\Ppi         {\ensuremath{\uppi}\xspace}

 \def\PDelta      {\ensuremath{\Delta}\xspace}                 
 \def\PXi      {\ensuremath{\Xi}\xspace}                 
 \def\PLambda      {\ensuremath{\Lambda}\xspace}                 
 \def\PSigma      {\ensuremath{\Sigma}\xspace}                 
 \def\POmega      {\ensuremath{\Omega}\xspace}                 
 \def\PUpsilon      {\ensuremath{\Upsilon}\xspace}                 
 
 \def\PB      {\ensuremath{\mathrm{B}}\xspace}                 
                  
 \def\PD      {\ensuremath{\mathrm{D}}\xspace}

 \def\PK      {\ensuremath{\mathrm{K}}\xspace}

 \def\PW      {\ensuremath{\mathrm{W}}\xspace}

 \def\Pb      {\ensuremath{\mathrm{b}}\xspace}                 
 \def\Pc      {\ensuremath{\mathrm{c}}\xspace}

 \def\Pi      {\ensuremath{\mathrm{i}}\xspace}

 \def\Pp      {\ensuremath{\mathrm{p}}\xspace}

 \def\Ps      {\ensuremath{\mathrm{s}}\xspace}

}
{

 \def\Ppi         {\ensuremath{\pi}\xspace}

 \mathchardef\PDelta="7101
 \mathchardef\PXi="7104
 \mathchardef\PLambda="7103
 \mathchardef\PSigma="7106
 \mathchardef\POmega="710A
 \mathchardef\PUpsilon="7107
                  
 \def\PB      {\ensuremath{B}\xspace}                 
                  
 \def\PD      {\ensuremath{D}\xspace}

 \def\PK      {\ensuremath{K}\xspace}

 \def\PW      {\ensuremath{W}\xspace}

 \def\Pb      {\ensuremath{b}\xspace}                 
 \def\Pc      {\ensuremath{c}\xspace}

 \def\Pi      {\ensuremath{i}\xspace}

 \def\Pp      {\ensuremath{p}\xspace}

 \def\Ps      {\ensuremath{s}\xspace}

}

\makeatletter
\ifcase \@ptsize \relax
  \newcommand{\miniscule}{\@setfontsize\miniscule{4}{5}}
\or
  \newcommand{\miniscule}{\@setfontsize\miniscule{5}{6}}
\or
  \newcommand{\miniscule}{\@setfontsize\miniscule{5}{6}}
\fi
\makeatother

\DeclareRobustCommand{\optbar}[1]{\shortstack{{\miniscule (\rule[.5ex]{1.25em}{.18mm})}
  \\ [-.7ex] $#1$}}

\def\W      {{\ensuremath{\PW}}\xspace}

\def\squark    {{\ensuremath{\Ps}}\xspace}

\def\cquark    {{\ensuremath{\Pc}}\xspace}

\def\bquark    {{\ensuremath{\Pb}}\xspace}

\def\pion   {{\ensuremath{\Ppi}}\xspace}

\def\pip    {{\ensuremath{\pion^+}}\xspace}
\def\pim    {{\ensuremath{\pion^-}}\xspace}
\def\pipm   {{\ensuremath{\pion^\pm}}\xspace}

\def\kaon    {{\ensuremath{\PK}}\xspace}
  \def\Kbar    {{\kern 0.2em\overline{\kern -0.2em \PK}{}}\xspace}

\def\KorKbar    {\kern 0.18em\optbar{\kern -0.18em K}{}\xspace}

\def\Kp      {{\ensuremath{\kaon^+}}\xspace}
\def\Km      {{\ensuremath{\kaon^-}}\xspace}
\def\Kpm     {{\ensuremath{\kaon^\pm}}\xspace}

  \def\Dbar    {{\kern 0.2em\overline{\kern -0.2em \PD}{}}\xspace}
\def\D       {{\ensuremath{\PD}}\xspace}

\def\DorDbar    {\kern 0.18em\optbar{\kern -0.18em D}{}\xspace}
\def\Dz      {{\ensuremath{\D^0}}\xspace}

\def\Dp      {{\ensuremath{\D^+}}\xspace}
\def\Dm      {{\ensuremath{\D^-}}\xspace}

\def\Dstar   {{\ensuremath{\D^*}}\xspace}

\def\Dstarp  {{\ensuremath{\D^{*+}}}\xspace}

\def\Ds      {{\ensuremath{\D^+_\squark}}\xspace}

\def\Dsm     {{\ensuremath{\D^-_\squark}}\xspace}
\def\Dspm    {{\ensuremath{\D^{\pm}_\squark}}\xspace}
\def\Dsmp    {{\ensuremath{\D^{\mp}_\squark}}\xspace}

\def\B       {{\ensuremath{\PB}}\xspace}
\def\Bbar    {{\ensuremath{\kern 0.18em\overline{\kern -0.18em \PB}{}}}\xspace}

\def\BorBbar    {\kern 0.18em\optbar{\kern -0.18em B}{}\xspace}
\def\Bz      {{\ensuremath{\B^0}}\xspace}

\def\Bs      {{\ensuremath{\B^0_\squark}}\xspace}

  \def\Y#1S{\ensuremath{\PUpsilon{(#1S)}}\xspace}

\def\proton      {{\ensuremath{\Pp}}\xspace}

\def\Lz          {{\ensuremath{\PLambda}}\xspace}
\def\Lbar        {{\ensuremath{\kern 0.1em\overline{\kern -0.1em\PLambda}}}\xspace}
\def\LorLbar    {\kern 0.18em\optbar{\kern -0.18em \PLambda}{}\xspace}

\def\Lb      {{\ensuremath{\Lz^0_\bquark}}\xspace}

\def\Lc      {{\ensuremath{\Lz^+_\cquark}}\xspace}

\def\BF         {{\ensuremath{\cal B}}\xspace}

\def\BR         {\BF}
\newcommand{\decay}[2]{\ensuremath{#1\!\to #2}\xspace}         

\def\to                 {\ensuremath{\rightarrow}\xspace}

\def\CP                {{\ensuremath{C\!P}}\xspace}

\def\AT#1     {\ensuremath{A_{\mathrm{T}}^{#1}}\xspace}           

\def\C#1      {\ensuremath{\mathcal{C}_{#1}}\xspace}                       
\def\Cp#1     {\ensuremath{\mathcal{C}_{#1}^{'}}\xspace}                    
\def\Ceff#1   {\ensuremath{\mathcal{C}_{#1}^{\mathrm{(eff)}}}\xspace}        
\def\Cpeff#1  {\ensuremath{\mathcal{C}_{#1}^{'\mathrm{(eff)}}}\xspace}       
\def\Ope#1    {\ensuremath{\mathcal{O}_{#1}}\xspace}                       
\def\Opep#1   {\ensuremath{\mathcal{O}_{#1}^{'}}\xspace}                    

\newcommand{\tev}{\ifthenelse{\boolean{inbibliography}}{\ensuremath{~T\kern -0.05em eV}\xspace}{\ensuremath{\mathrm{\,Te\kern -0.1em V}}}\xspace}
\newcommand{\gev}{\ensuremath{\mathrm{\,Ge\kern -0.1em V}}\xspace}
\newcommand{\mev}{\ensuremath{\mathrm{\,Me\kern -0.1em V}}\xspace}
\newcommand{\kev}{\ensuremath{\mathrm{\,ke\kern -0.1em V}}\xspace}
\newcommand{\ev}{\ensuremath{\mathrm{\,e\kern -0.1em V}}\xspace}
\newcommand{\gevc}{\ensuremath{{\mathrm{\,Ge\kern -0.1em V\!/}c}}\xspace}
\newcommand{\mevc}{\ensuremath{{\mathrm{\,Me\kern -0.1em V\!/}c}}\xspace}
\newcommand{\gevcc}{\ensuremath{{\mathrm{\,Ge\kern -0.1em V\!/}c^2}}\xspace}
\newcommand{\gevgevcccc}{\ensuremath{{\mathrm{\,Ge\kern -0.1em V^2\!/}c^4}}\xspace}
\newcommand{\mevcc}{\ensuremath{{\mathrm{\,Me\kern -0.1em V\!/}c^2}}\xspace}

\def\mum  {\ensuremath{{\,\upmu\rm m}}\xspace}

\def\invpb {\ensuremath{\mbox{\,pb}^{-1}}\xspace}

\def\invfb   {\ensuremath{\mbox{\,fb}^{-1}}\xspace}

\newcommand{\stat}{\ensuremath{\mathrm{\,(stat)}}\xspace}
\newcommand{\syst}{\ensuremath{\mathrm{\,(syst)}}\xspace}

\newcommand{\chisq}{\ensuremath{\chi^2}\xspace}

\def\gsim{{~\raise.15em\hbox{$>$}\kern-.85em
          \lower.35em\hbox{$\sim$}~}\xspace}
\def\lsim{{~\raise.15em\hbox{$<$}\kern-.85em
          \lower.35em\hbox{$\sim$}~}\xspace}

\def\sPlot{\mbox{\em sPlot}\xspace}

\def\sqs   {\ensuremath{\protect\sqrt{s}}\xspace}

\def\ptot       {\mbox{$p$}\xspace}
\def\pt         {\mbox{$p_{\rm T}$}\xspace}

\def\evtgen     {\mbox{\textsc{EvtGen}}\xspace}

\def\geant      {\mbox{\textsc{Geant4}}\xspace}

\def\pythia     {\mbox{\textsc{Pythia}}\xspace}

\def\tell1  {TELL1\xspace}
\def\ukl1   {UKL1\xspace}

\newcommand{\mytev}{\ensuremath{\mathrm{\,Te\kern -0.1em V}}\xspace} 

\newcommand{\0}{\hphantom{0}}
\newcommand{\pdecay}[2]{\ensuremath{#1(\!\to #2)}\xspace} 

\newcommand{\DtoKpipi}{\decay{\Dm}{\Kp\pim\pim}}
\newcommand{\DstoKKpi}{\decay{\Dsm}{\Kp\Km\pim}}

\newcommand{\BorBs}{\ensuremath{\B^{0}_{(\squark)}}\xspace}

\newcommand{\DorDs}{\ensuremath{\D^{\pm}_{(\squark)}}\xspace}

\newcommand{\BdtoDpi}{\decay{\Bz}{\Dm\pip}}

\newcommand{\BdtoDsK}{\decay{\Bz}{\Dsm\Kp}}

\newcommand{\BstoDsK}{\decay{\Bs}{\Dsmp\Kpm}}

\newcommand{\BstoDspi}{\decay{\Bs}{\Dsm\pip}}

\newcommand{\BstoKKpipi}{\decay{\Bs}{\Kp\Km\pip\pim}}

\newcommand{\LbtoLcpi}{\decay{\Lb}{\Lc\pim}}

\newcommand{\LbtoDsp}{\decay{\Lb}{\Dsm\proton}}

\newcommand{\BdDPi}{\BdtoDpi}

\newcommand{\BdDsK}{\BdtoDsK}
\newcommand{\BsDsPi}{\BstoDspi}
\newcommand{\BsDsK}{\BstoDsK}

\usepackage{longtable}
\usepackage{cite}
\usepackage{mciteplus}
\mciteErrorOnUnknownfalse

\begin{document}
\renewcommand{\thefootnote}{\fnsymbol{footnote}}
\setcounter{footnote}{1}

\begin{titlepage}
\pagenumbering{roman}

\vspace*{-1.5cm}
\centerline{\large EUROPEAN ORGANIZATION FOR NUCLEAR RESEARCH (CERN)}
\vspace*{1.5cm}
\hspace*{-0.5cm}
\begin{tabular*}{\linewidth}{lc@{\extracolsep{\fill}}r}
\ifthenelse{\boolean{pdflatex}}
{\vspace*{-2.7cm}\mbox{\!\!\!\includegraphics[width=.14\textwidth]{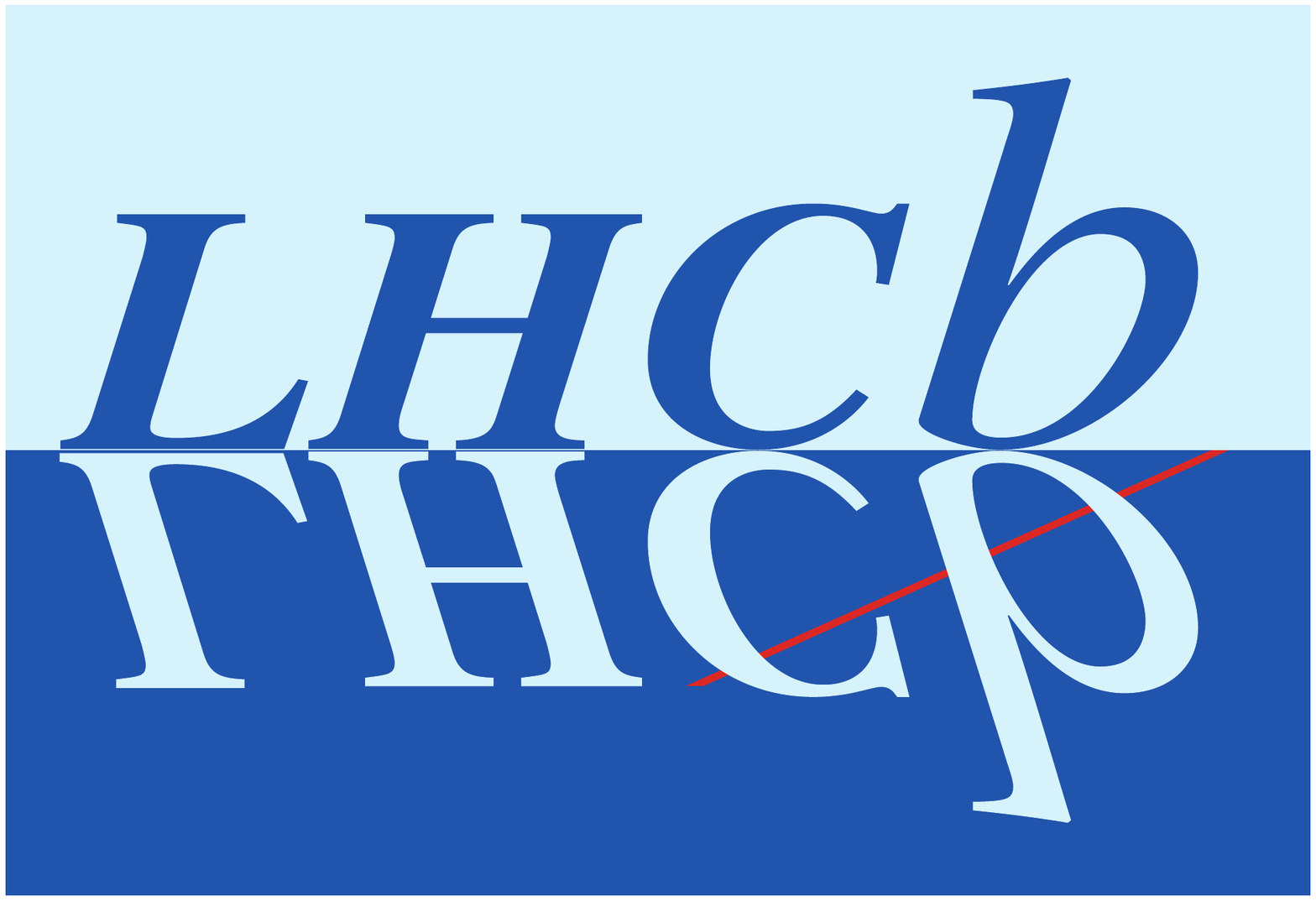}} & &}%
{\vspace*{-1.2cm}\mbox{\!\!\!\includegraphics[width=.12\textwidth]{lhcb-logo.eps}} & &}%
\\
 & & CERN-PH-EP-2014-297 \\ 
 & & LHCb-PAPER-2014-064 \\
 & & 23 December 2014 \\
 & & \\
\end{tabular*}

\vspace*{1.0cm}

{\bf\boldmath\huge
\begin{center}
    Determination of the branching fractions of \BsDsK~and \BdDsK
\end{center}
}

\vspace*{2.0cm}

\begin{center}
The LHCb collaboration\footnote{Authors are listed at the end of this paper.}
\end{center}

\vspace{\fill}

\begin{abstract}
  \noindent
  Measurements are presented of the branching fractions of the decays
  \BsDsK~and \BdDsK relative to the decays \BsDsPi~and \BdDPi, respectively.
  The data used correspond to an integrated luminosity of~$3.0\invfb$ of
  proton-proton collisions. The ratios of branching fractions are
  \[ \dfrac{\BF(\BsDsK)}{\BF(\BsDsPi)} =\, 0.0752 \pm 0.0015 \pm 0.0019  \]
  and
  \[ \dfrac{\BF(\BdDsK)}{\BF(\BdDPi)}  =\, 0.0129 \pm 0.0005 \pm 0.0008, \]
  where the uncertainties are statistical and systematic, respectively.
\end{abstract}

\vspace*{2.0cm}

\begin{center}
  Submitted to JHEP
\end{center}

\vspace{\fill}

{\footnotesize 
\centerline{\copyright~CERN on behalf of the \lhcb collaboration, license \href{http://creativecommons.org/licenses/by/4.0/}{CC-BY-4.0}.}}
\vspace*{2mm}

\end{titlepage}

\newpage
\setcounter{page}{2}
\mbox{~}

\cleardoublepage

\renewcommand{\thefootnote}{\arabic{footnote}}
\setcounter{footnote}{0}

\pagestyle{plain}
\setcounter{page}{1}
\pagenumbering{arabic}

\section{Introduction}
\label{sec:Introduction}

\begin{figure}[b]
  \begin{center}
    \includegraphics[width=.32\linewidth]{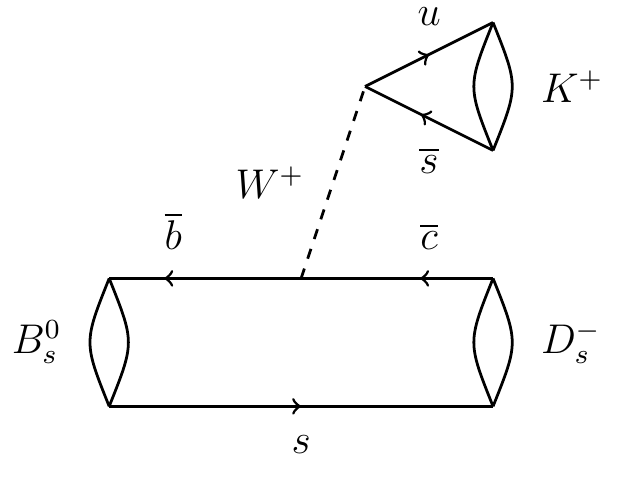}
    \hspace{2em}
    \includegraphics[width=.32\linewidth]{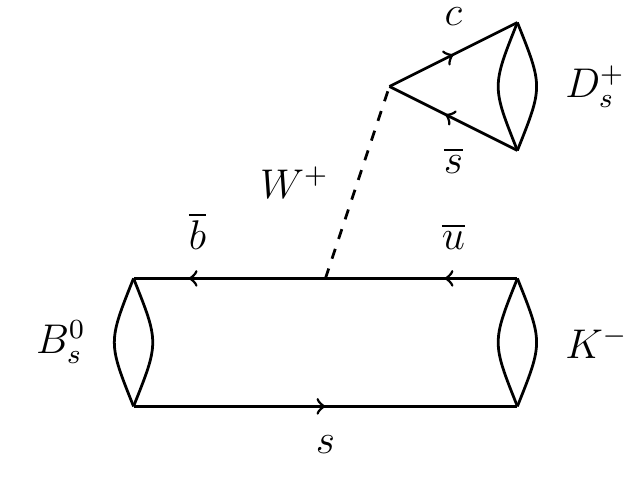}
    \includegraphics[width=.32\linewidth]{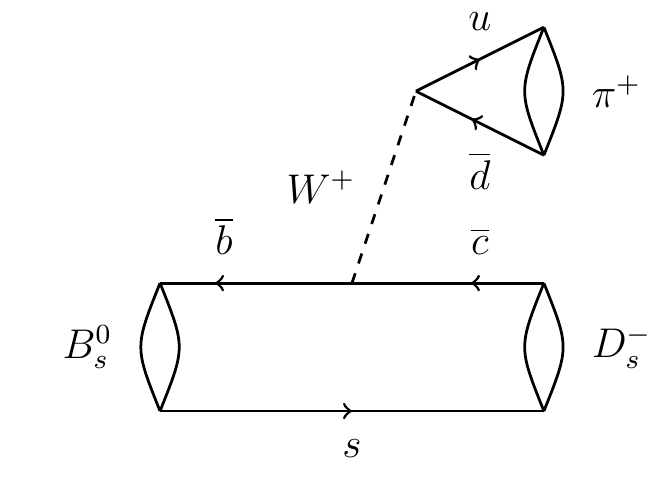}
    \hspace{3.5em}
    \includegraphics[width=.32\linewidth]{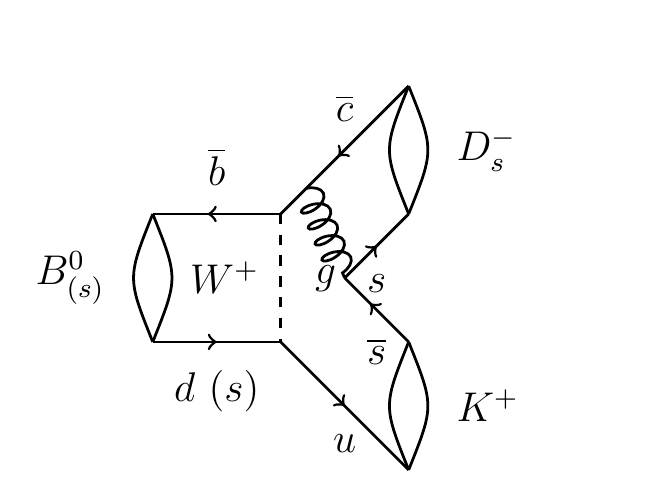}
  \end{center}
  \caption{
  \small Feynman diagrams of the processes under study. The upper diagrams
  represent the two tree topologies in which a \Bs~meson decays into the
  $\Dsmp\Kpm$~final state, and the lower diagrams show the tree diagram of
  \BsDsPi~and the \W-exchange topology of $\B^0_{(s)}\to\Dsm\Kp$.}
  \label{fig:FeynmanDiagrams}
\end{figure}

The decays \BsDsK offer a prime opportunity to measure \CP~violation in the
interference between mixing and decay~\cite{Aleksan:1991nh,Fleischer:2003}.  The
\Bs~meson can decay into both charge-conjugate decays, providing sensitivity to
the CKM angle $\gamma$~\cite{LHCb-PAPER-2014-038}.  The decays \BsDsK and
\BsDsPi occur predominantly through colour-allowed tree diagrams (see
Fig.~\ref{fig:FeynmanDiagrams}).  
A lower bound on the ratio of the \BsDsK and \BsDsPi branching fractions was
derived, $\BF(\BsDsK)/\BF(\BsDsPi) \geq 0.080 \pm 0.007$~\cite{DeBruyn:2012},
with minimal external experimental and theoretical input.  
Using SU(3) flavour symmetry, and measurements of \BdDPi~decays at the
\B-factories, a prediction for the ratio of branching fractions was calculated,
$\BF(\BsDsK)/\BF(\BsDsPi) = 0.086^{+0.009}_{-0.007}$~\cite{DeBruyn:2012}, where
the uncertainty includes contributions from non-factorisable
effects~\cite{Fleischer:2010ca} and from possible SU(3)-breaking effects of up
to 20\%.  Contributions from the $W$-exchange diagram, absent in the decay
\BsDsPi, were estimated from the \BdDsK branching fraction.

The \cdf and \belle collaborations have pioneered the study of this
ratio~\cite{Aaltonen:2008ab,Louvot:2008sc}, followed by the \lhcb collaboration,
which measured a ratio lower than the theoretical
bound~\cite{LHCb-PAPER-2011-022}, using data corresponding to an integrated
luminosity of $336\invpb$.  This paper presents an update for the absolute
branching fraction of \BsDsK, and for the branching fraction of \BsDsK relative
to that of \BsDsPi.  

In addition, the ratio of branching fractions of the \BdDsK
and \BdDPi decays is reported.  The decay \BdDsK proceeds through the
colour-suppressed $W$-exchange diagram and the branching ratio determination
allows the size of the $W$-exchange amplitude to be estimated, for example in the
\BsDsK decay.  The existing branching fraction measurements by BaBar and Belle,
$\BR(\BdDsK) = (2.9 \pm 0.4\stat \pm 0.2\syst)\times 10^{-5}$~\cite{Aubert:2008zi}
and $(1.91 \pm 0.24\stat \pm 0.17\syst)\times 10^{-5}$~\cite{Das:2010be},
respectively, show a difference of about 1.8 standard deviations, 
and suggest that an enhancement of the branching fraction due to rescattering effects is
small~\cite{Gronau:2008kk}.
Note that throughout this paper, charge conjugation is implied, and thus that the
branching fraction $\BR(B^0_{(s)})$ corresponds to the average of the branching
fraction of the $B^0_{(s)}$~decay and the 
$\kern 0.18em\overline{\kern -0.18em B}^0_{(s)}$~decay.

The $\proton\proton$-collision data used in this analysis correspond to an
integrated luminosity of $3.0\invfb$, of which $1.0\invfb$ was collected by
\lhcb in 2011 at a centre-of-mass energy of $\sqs=7\tev$, and the remaining
$2.0\invfb$ in 2012 at $\sqs=8\tev$.

The \lhcb detector~\cite{Alves:2008zz} is a single-arm forward spectrometer
covering the \mbox{pseudorapidity} range $2<\eta <5$, designed for the study of
particles containing \bquark or \cquark quarks. The detector includes a
high-precision tracking system consisting of a silicon-strip vertex detector
surrounding the $pp$ interaction region, a large-area silicon-strip detector
located upstream of a dipole magnet with a bending power of about $4{\rm\,Tm}$,
and three stations of silicon-strip detectors and straw drift tubes placed
downstream of the magnet. The polarity of the magnet is reversed periodically
throughout data-taking. The tracking system provides a measurement of momentum,
\ptot,  with a relative uncertainty that varies from 0.4\% at low momentum to
0.6\% at 100\gevc.  The minimum distance of a track to a primary vertex, the
impact parameter, is measured with a resolution of $(15+29/\pt)\mum$, where \pt
is the component of \ptot transverse to the beam, in \gevc.  Different types of
charged hadrons are distinguished using information from two ring-imaging
detectors.

The trigger consists of a hardware stage, based on information from the
calorimeter and muon systems, followed by a software stage, which applies a
full event reconstruction.  The software trigger requires a two-, three- or
four-track secondary vertex with a significant displacement from the primary
$pp$ interaction vertices~(PVs). At least one charged particle must have a
transverse momentum $\pt > 1.7\gevc$ and be inconsistent with originating from
the PV. A multivariate algorithm~\cite{BBDT} is used for the identification of
secondary vertices consistent with the decay of a \bquark hadron.

In the simulation, $pp$ collisions are generated using
\pythia~\cite{Sjostrand:2006za,*Sjostrand:2007gs} with a specific \lhcb
configuration~\cite{LHCb-PROC-2010-056}. Decays of hadronic particles are
described by \evtgen~\cite{Lange:2001uf}. The interaction of the generated
particles with the detector and its response are implemented using the \geant
toolkit~\cite{Allison:2006ve, *Agostinelli:2002hh} as described in
Ref.~\cite{LHCb-PROC-2011-006}.

\section{Event selection}

Candidate \BorBs~mesons are reconstructed by combining a \DorDs~candidate
decaying into three light hadrons, \DtoKpipi or \DstoKKpi, with an additional
pion or kaon (the ``bachelor'' particle). Each of the four final-state light
hadrons is required to have a good track quality, high momentum and transverse
momentum, and a large impact parameter with respect to the primary vertex. The
contribution from charmless \BorBs~decays, such as \BstoKKpipi, is suppressed
by requiring the \DorDs~candidate to have a significant flight distance from
the reconstructed \BorBs~decay vertex, and by requiring its mass to fall within
a small mass window of ${}^{+22}_{-24}\mevcc$~around the
\DorDs~mass~\cite{PDG2014}. To reduce the combinatorial background, a
multivariate algorithm is applied. This boosted decision tree
(BDT)~\cite{Breiman,AdaBoost} is identical to that used in the analysis of the
\CP~asymmetry in \BsDsK~decays~\cite{LHCb-PAPER-2014-038}, and was trained with
\BsDsPi candidates from data, using a weighted data sample based on the \sPlot
technique~\cite{Pivk:2004ty} as signal and candidates with an invariant mass
greater than $5445\mevcc$~as background. The variables with the highest
discriminating power are found to be the difference between the \chisq~from the
vertex fit of the associated PV reconstructed with and without the considered
\bquark-hadron candidate, the \pt~of the final-state particles, and the angle
between the \bquark-hadron momentum vector and the vector connecting its
production and decay vertices.

Misidentification of particles leads to peaking backgrounds in the signal
region, for example \BsDsPi~events reconstructed as \BsDsK~candidates.
Pions and kaons in these decays are required to satisfy particle identification
(PID) requirements, and approximately $60\%$ of the signal is retained while
over $99\%$ of the background is rejected. The efficiencies of these requirements
are determined by studying kinematically selected
$\decay{\Dstarp}{\pdecay{\Dz}{\Km\pip}\pip}$ and $\decay{\Lz}{\proton\pim}$
decays obtained from data, which provide high-purity PID.

\begin{table}[!t]
    \caption{\small{Kinematic and PID selection efficiencies for each signal
            decay, as determined from simulated events and data, respectively.
            The kinematic efficiencies represent weighted averages determined
            from events simulated at $\sqs = 7\mytev$ ($34\%$) and $\sqs =
            8\mytev$ ($66\%$). The binomial uncertainties result from the size
    of the simulated samples.}}
    \begin{center}\begin{tabular}{lccc}
        \hline
        Selection efficiency (\%) & Kinematic & PID                 & Total          \\
        \hline
        \BdtoDpi  & $1.89 \pm 0.01$           & $74.29 \pm 0.07$    & $1.40 \pm 0.01$\\
        \BstoDspi & $1.92 \pm 0.02$           & $67.10 \pm 0.09$    & $1.29 \pm 0.01$\\
        \BstoDsK  & $2.08 \pm 0.01$           & $55.52 \pm 0.17$    & $1.15 \pm 0.01$\\
        \BdtoDsK  & $1.70 \pm 0.03$           & $58.11 \pm 0.82$    & $0.99 \pm 0.02$\\
        \hline
    \end{tabular}\end{center}
\label{tab:Efficiencies}
\end{table}

In the \BdDPi selection, loose PID requirements are applied since the branching
fraction of the signal process is much larger than those of background decays
resulting from misidentification. In the \BsDsPi and \BsDsK selections, a
stricter requirement is applied to the \Ds~decays to distinguish \Ds~and
\Dp~mesons. For these decays, a further selection requirement is applied to
reduce the background from \LbtoLcpi decays, where one of the
\Ds~daughters is misidentified as a proton. This requirement removes any candidate
which fulfils two criteria: that there is a large probability for one of the
\Ds~daughters to be a misidentified proton, and that when the \Ds~decay is
reconstructed under the \Lc~hypothesis, its invariant mass falls within
$21\mevcc$ of the nominal \Lc~mass~\cite{PDG2014}.  This procedure almost fully
eliminates this background. The efficiency of the selection is obtained from
simulation and is summarised in Table~\ref{tab:Efficiencies}.

\section{Signal yield determination}

An unbinned maximum likelihood fit to the candidate invariant mass distribution
is performed for each of the three final states. The signal shapes are
parametrised by a double-sided Crystal Ball shape~\cite{Skwarnicki:1986xj}. This
function consists of a central Gaussian part, whose mean and width are free
parameters, and power-law tails on both lower and upper sides, to account for
energy loss due to final-state radiation and detector resolution effects. The
functional form for the combinatorial background, an exponential function with
an offset, is obtained from same-charge $\Dspm\pipm$~combinations. All
parameters of the combinatorial background are left free in the fit to data.

\begin{figure}[tb]
  \begin{center}
    \includegraphics[width=.8\linewidth]{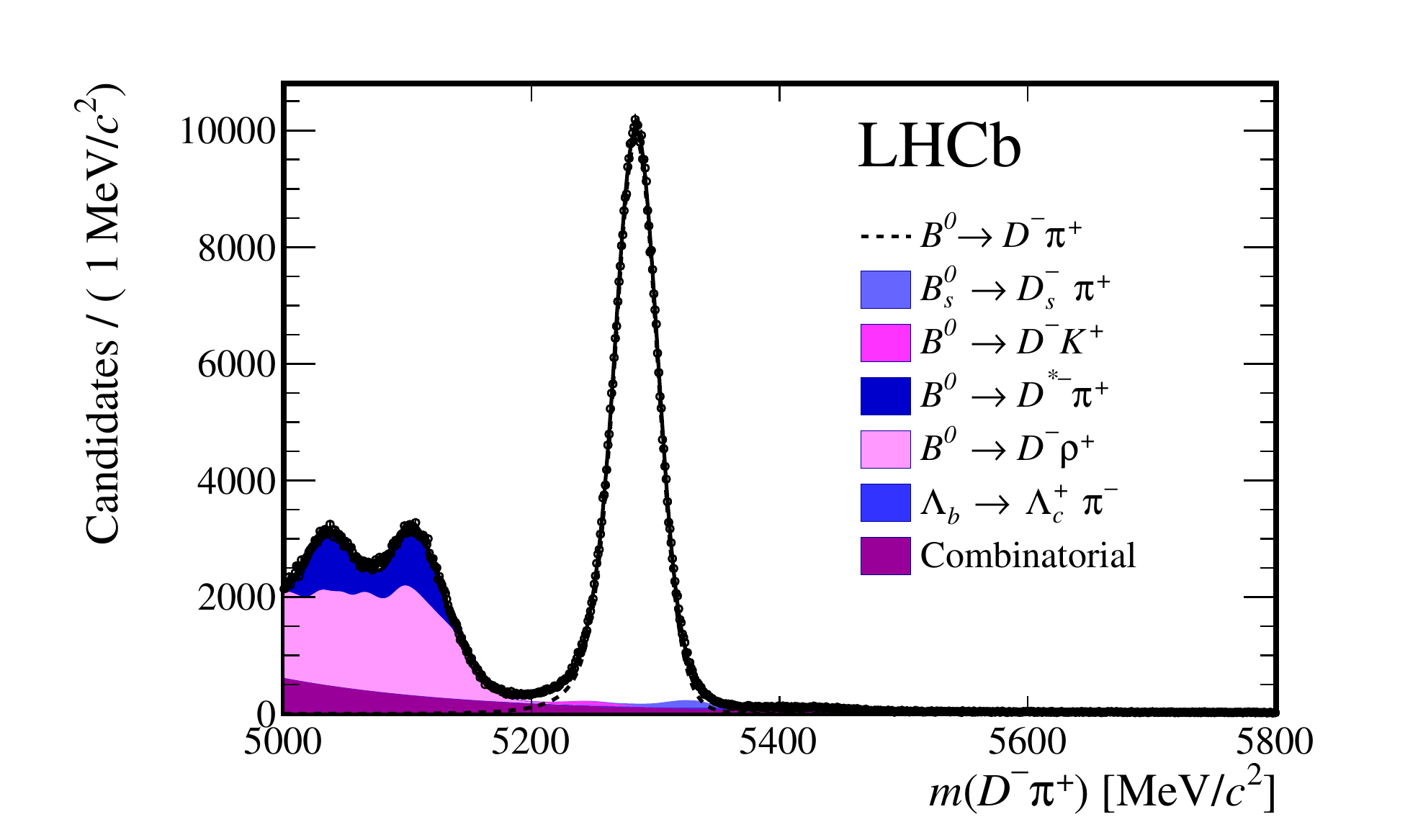} \put(-260,150){(a)}\\
    \includegraphics[width=.8\linewidth]{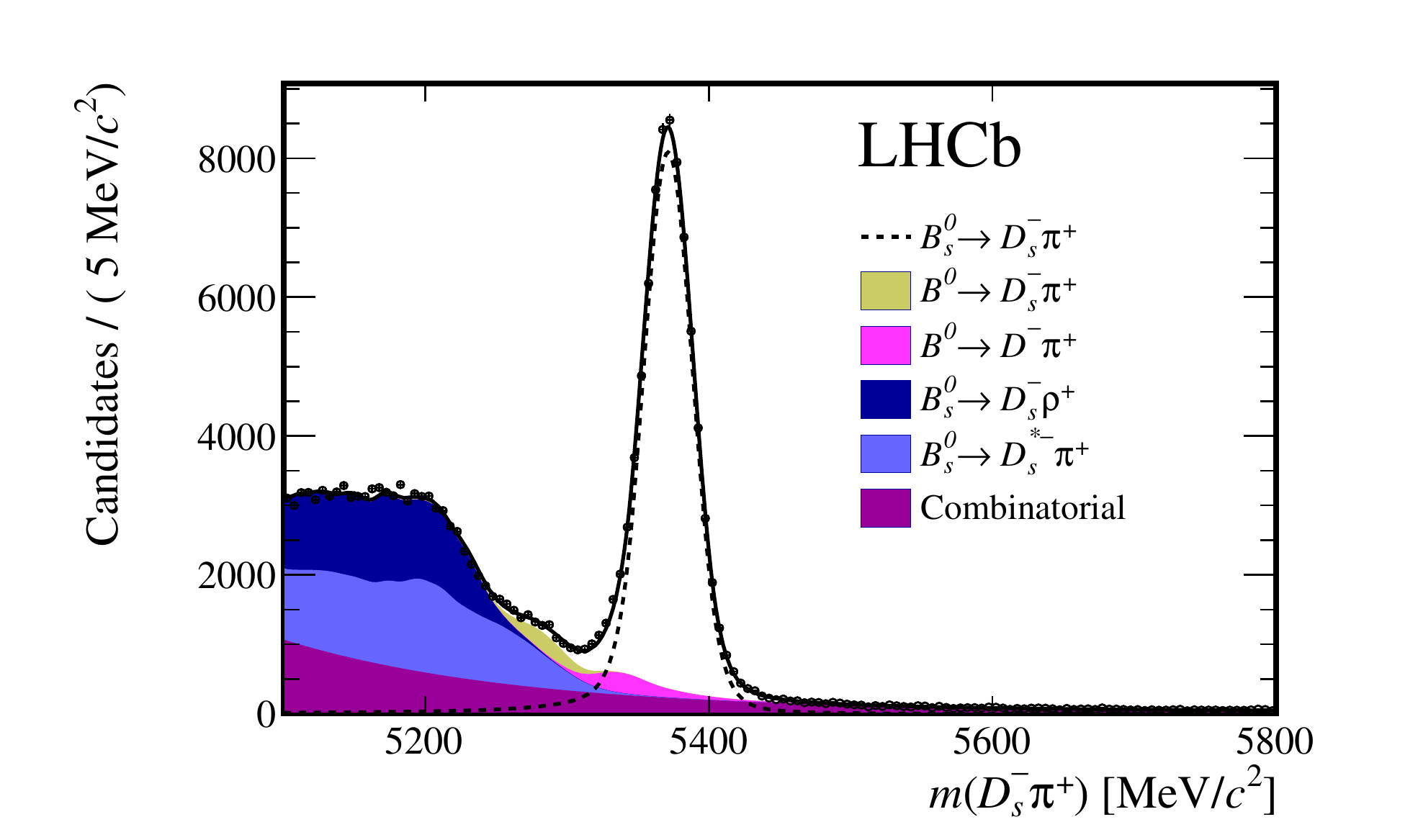}\put(-260,150){(b)}
  \end{center}
  \caption{
    \small Results of the fits to the invariant mass distributions of the final states (a)~$\Dm\pip$ and
           (b)~$\Dsm\pip$.}
  \label{fig:FitResults1}
\end{figure}

\begin{figure}[tb]
  \begin{center}
    \includegraphics[width=0.9\linewidth]{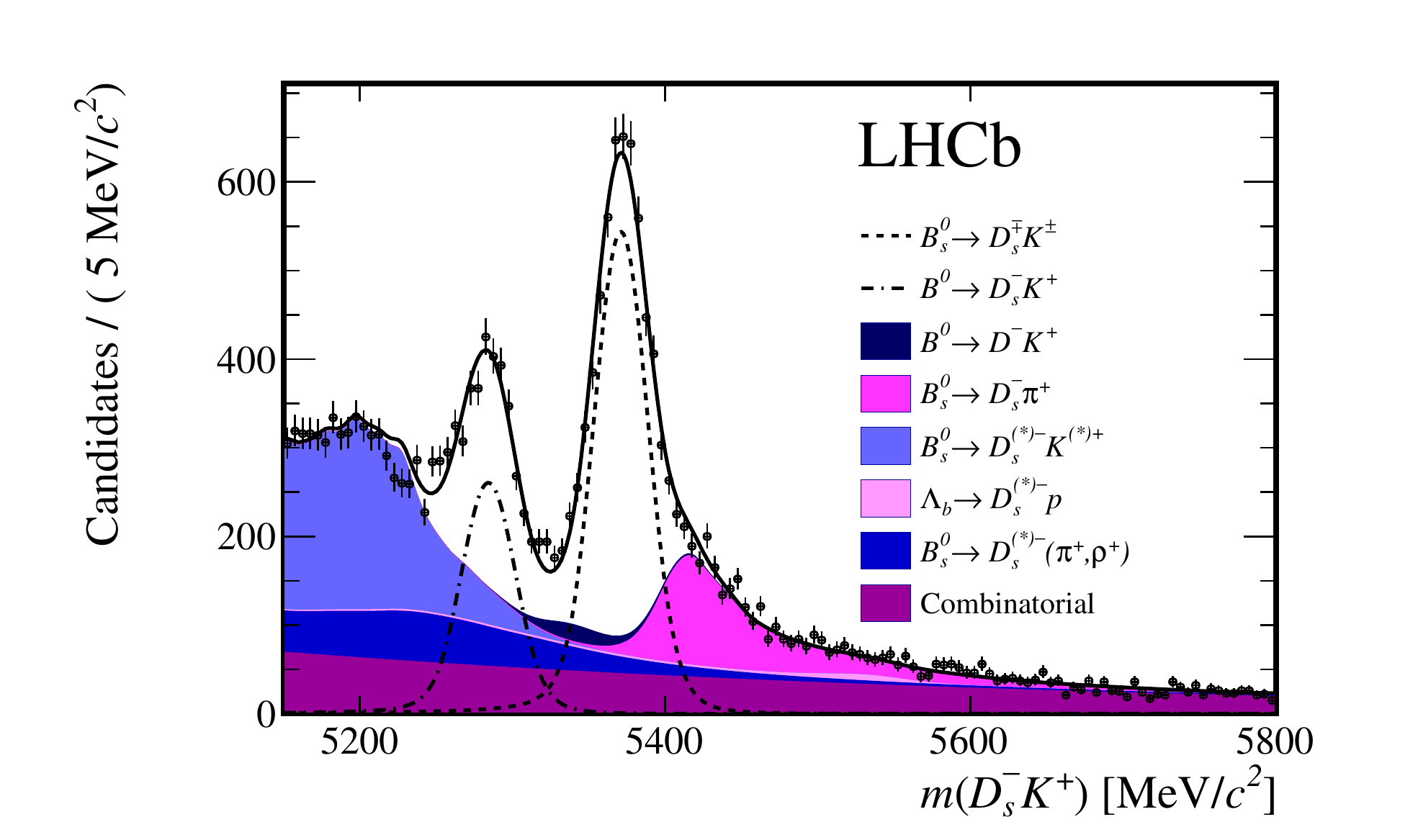}
  \end{center}
  \caption{
    \small Result of the fit to the invariant mass distribution of the final state $\Dsmp\Kpm$.}
  \label{fig:FitResults2}
\end{figure}

The physical backgrounds can be split into two categories: misidentified
backgrounds, predominantly where one of the final state pions (kaons) is mistaken for a kaon
(pion); and partially reconstructed backgrounds, where a neutral pion or a
photon is not included in the candidate reconstruction, causing the
reconstructed \BorBs~mass to shift to lower values. Some backgrounds fall into
both categories. The number of background components considered varies per
final state. The invariant mass shapes of these backgrounds are obtained from
simulation at $\sqs = 8\tev$, with the event selection applied.  The yield of
each background is a parameter in the fit, with most background components
Gaussian-constrained around the expected yield normalised to the \BdDPi~yield obtained from data. 
The constraints are
assigned an uncertainty of $10\%$, which reflects the uncertainties from
production fractions, branching fractions, and reconstruction efficiencies. 
The resulting background yields from the fit are close to the expected values.  
The results of the fits for the three final states are shown in
Figs.~\ref{fig:FitResults1} and~\ref{fig:FitResults2}, and in
Table~\ref{tab:Yields}. The three fits are independent, and no parameters are
shared among them.

\begin{table}[!b]
    \caption{\small{Yields for the four signal decay types, as
    obtained from the fits.}}
    \begin{center}\begin{tabular}{lr}
        \hline
                  & Yield            \\
        \hline
        \BdtoDpi  & $458\,940\pm 959$\\
        \BstoDspi & $ 75\,566\pm 342$\\
        \BstoDsK  & $  5\,101\pm 100$\\
        \BdtoDsK  & $  2\,452\pm\098$\\
        \hline
    \end{tabular}\end{center}
\label{tab:Yields}
\end{table}

Various consistency checks are performed for each of the fits. The fitted yield
of \BsDsPi~events reconstructed in the \Dsmp\Kpm~final state, which is allowed
to vary in the fit, is consistent with the expected yield based on the relative
branching fraction, particle misidentification probability and reconstruction
efficiency. For each of the fits, consistency is also found between the fitted
yield for both magnet polarities separately and the fraction of data
corresponding to that polarity. This demonstrates that the relative yields are
stable as a function of time and magnet polarity.

\section{Systematic uncertainties}

Systematic uncertainties arise from the fit model and the candidate selection,
and are summarised in Table~\ref{tab:Systematics}. The systematic uncertainty
from the fit model is determined by applying variations to the fit model and
comparing the yield to the nominal result, taking the difference as a
systematic uncertainty. These variations include a different combinatorial
shape, fixing the signal shape tail parameters to values obtained from
simulation, and using background shapes determined from simulation matching the
\lhcb conditions during 2011 ($\sqs = 7\tev$). In the $\Dm\pip$~analysis, the
fit range is reduced to start at $5100\mevcc$. In the $\Dsmp\Kpm$~analysis, the
as yet unobserved decay \LbtoDsp~is omitted from the fit.

The uncertainty on the candidate selection is separated into three parts: the
uncertainty due to the differences between data and simulation, that due to the
PID requirements on the final state pions and kaons, and that due to the
hardware trigger efficiency. The first of these uncertainties is determined
from the selection efficiency difference between magnet polarities in
simulation, and by estimating the uncertainty on the BDT selection efficiency
due to differences between data and simulation. This is calculated by
reweighting simulated events to match the data more closely, and calculating
the difference in BDT efficiency between those and the unweighted samples.
The uncertainty on the PID efficiency and misidentification rate is estimated
by comparing the PID performance measured using a simulated \Dstar~calibration
sample with that observed in simulated signal events.
The systematic uncertainty from the hardware trigger efficiency arises from
differences in the pion and kaon trigger efficiencies which are not reproduced
in the simulation. The uncertainty is scaled with the fraction of events where
a signal track was responsible for triggering.

A further systematic uncertainty is added to account for possible charmless
\Bz~decays peaking under the \BdDsK~signal.  Some of the uncertainties cancel
in the ratios of branching fractions, leading to lower overall systematic
uncertainties than those determined individually for each decay channel.

\begin{table}[!t]
    \caption{\small{Systematic uncertainties on the ratios of branching
    fractions, in $\%$, obtained as described in the text. The total
    uncertainty is obtained by adding the separate contributions in
    quadrature.}}
    \begin{center}\begin{tabular}{lcc}
        \hline\vspace{-0.7em}\\
        Source                                &  $\dfrac{\BstoDsK}{\BstoDspi}$ & $\dfrac{\BdtoDsK}{\BdtoDpi}$ \vspace{0.3em}\\
        \hline
        Fit model                             &  $1.1$                         & $3.6$                                      \\
        Candidate selection                   &  $2.1$                         & $2.9$                                      \\
        Hardware trigger                      &  $1.0$                         & $1.2$                                      \\
        Charmless background                  &  --                            & $1.0$                                      \\
        \hline
        Total                                 &  $2.5$                         & $4.9$                                      \\
        \hline
    \end{tabular}\end{center}
\label{tab:Systematics}
\end{table}

\section{Determination of branching fractions}

The ratios of branching fractions are evaluated using the expression
\begin{equation}
\dfrac{\BF(A)}{\BF(B)} =
    \dfrac{\varepsilon_{B}}{\varepsilon_{A}}
    \dfrac{N_{A}}{N_{B}}
    \dfrac{f_{B}}{f_{A}}
    \dfrac{\BF_{\DorDs}}{\BF_{\DorDs}},
\label{eqn:BF}
\end{equation}
where $\varepsilon_{X}$, $f_{X}$ and $N_{X}$ are the selection efficiency, the
hadronisation fraction, and the fitted yield of decay $X$, respectively, and
$\BF_{\DorDs}$ is the branching fraction of \DorDs~decays, as appropriate. The
following values are used as input~\cite{PDG2014}:
\begin{align}
\BF(\BdtoDpi)  =&\, (2.68 \pm 0.13) \times 10^{-3},\nonumber \\
\BF(\BstoDspi) =&\, (3.04 \pm 0.23) \times 10^{-3},\nonumber \\
\label{eqn:WorldAverageBranchingFractions}
\BF(\DtoKpipi) =&\, (9.13 \pm 0.19) \times 10^{-2},\nonumber \\
\BF(\DstoKKpi) =&\, (5.39 \pm 0.21) \times 10^{-2}.\nonumber 
\end{align}
As a cross-check, a value $\BF(\BsDsPi) = (2.95 \pm 0.01\stat) \times 10^{-3}$
was obtained from the measured \BsDsPi and \BdDPi yields using
Eq.~\eqref{eqn:BF}. This measurement is compatible with the world-average
value, and the central value is unchanged with respect to the
previous result published by LHCb~\cite{LHCb-PAPER-2011-022}.

The following results are obtained
\begin{align*}
\dfrac{\BF(\BsDsK)}{\BF(\BsDsPi)} =&\, 0.0752 \pm 0.0015\stat \pm 0.0019\syst, \\
       \BF(\BsDsK)                =&\, (2.29 \pm 0.05\stat \pm 0.06\syst \pm 0.17(\BF_{\Bs})) \times 10^{-4}, \\
\dfrac{\BF(\BdDsK)}{\BF(\BdDPi)}  =&\, 0.0129 \pm 0.0005\stat \pm 0.0007\syst \pm 0.0004(\BF_{\DorDs}), \\
       \BF(\BdDsK)                =&\, (3.45 \pm 0.14\stat \pm 0.20\syst \pm 0.20(\BF_{\Bz,\DorDs})) \times 10^{-5}, 
\end{align*}
where the uncertainties labelled $(\BF)$~arise from the uncertainties on the
branching fractions used as input.

The branching fractions of \BsDsK~and \BdDsK~presented here are more precise
than the current world-average values. The result for
$\BF(\BsDsK)/\BF(\BsDsPi)$ is compatible with theoretical
expectations~\cite{DeBruyn:2012}.  As expected~\cite{Fleischer:2010ca}, the
branching fraction of the decay \BdDsK, dominated by the \W-exchange topology,
is suppressed compared to the decay \BdDPi, which predominantly proceeds
through the colour-allowed tree topology.  The measured value of $\BF(\BdDsK)$
is in good agreement with existing measurements from the BaBar
collaboration~\cite{Aubert:2008zi}, and is larger than the result published by
the Belle collaboration~\cite{Das:2010be} with a significance of more than
three standard deviations.

\section*{Acknowledgements}

\noindent We express our gratitude to our colleagues in the CERN
accelerator departments for the excellent performance of the LHC. We
thank the technical and administrative staff at the LHCb
institutes. We acknowledge support from CERN and from the national
agencies: CAPES, CNPq, FAPERJ and FINEP (Brazil); NSFC (China);
CNRS/IN2P3 (France); BMBF, DFG, HGF and MPG (Germany); INFN (Italy); 
FOM and NWO (The Netherlands); MNiSW and NCN (Poland); MEN/IFA (Romania); 
MinES and FANO (Russia); MinECo (Spain); SNSF and SER (Switzerland); 
NASU (Ukraine); STFC (United Kingdom); NSF (USA).
The Tier1 computing centres are supported by IN2P3 (France), KIT and BMBF 
(Germany), INFN (Italy), NWO and SURF (The Netherlands), PIC (Spain), GridPP 
(United Kingdom).
We are indebted to the communities behind the multiple open 
source software packages on which we depend. We are also thankful for the 
computing resources and the access to software R\&D tools provided by Yandex LLC (Russia).
Individual groups or members have received support from 
EPLANET, Marie Sk\l{}odowska-Curie Actions and ERC (European Union), 
Conseil g\'{e}n\'{e}ral de Haute-Savoie, Labex ENIGMASS and OCEVU, 
R\'{e}gion Auvergne (France), RFBR (Russia), XuntaGal and GENCAT (Spain), Royal Society and Royal
Commission for the Exhibition of 1851 (United Kingdom).

\setboolean{inbibliography}{true}
\bibliographystyle{LHCb}
\bibliography{main,LHCb-PAPER,LHCb-CONF,LHCb-DP,LHCb-TDR}

\newpage

\centerline{\large\bf LHCb collaboration}
\begin{flushleft}
\small
R.~Aaij$^{41}$, 
B.~Adeva$^{37}$, 
M.~Adinolfi$^{46}$, 
A.~Affolder$^{52}$, 
Z.~Ajaltouni$^{5}$, 
S.~Akar$^{6}$, 
J.~Albrecht$^{9}$, 
F.~Alessio$^{38}$, 
M.~Alexander$^{51}$, 
S.~Ali$^{41}$
% ...

{\footnotesize \it
% ...
$ ^{5}$Clermont Universit\'{e}, Universit\'{e} Blaise Pascal, CNRS/IN2P3, LPC, Clermont-Ferrand, France\\
$ ^{6}$CPPM, Aix-Marseille Universit\'{e}, CNRS/IN2P3, Marseille, France\\
% ...
$ ^{9}$Fakult\"{a}t Physik, Technische Universit\"{a}t Dortmund, Dortmund, Germany\\
% ...
$ ^{37}$Universidad de Santiago de Compostela, Santiago de Compostela, Spain\\
$ ^{38}$European Organization for Nuclear Research (CERN), Geneva, Switzerland\\
% ...
$ ^{41}$Nikhef National Institute for Subatomic Physics, Amsterdam, The Netherlands\\
% ...
$ ^{46}$H.H. Wills Physics Laboratory, University of Bristol, Bristol, United Kingdom\\
% ...
$ ^{51}$School of Physics and Astronomy, University of Glasgow, Glasgow, United Kingdom\\
$ ^{52}$Oliver Lodge Laboratory, University of Liverpool, Liverpool, United Kingdom\\
% ...
}
\end{flushleft}

\end{document}